\RequirePackage{fix-cm}
\documentclass[twocolumn,epjc3]{svjour3} 
\usepackage{graphicx}
\usepackage{amssymb}
\usepackage{amsmath}
\usepackage{epstopdf}
\usepackage{array}
\newcolumntype{L}[1]{>{\raggedright\let\newline\\\arraybackslash\hspace{0pt}}m{#1}}
\newcolumntype{C}[1]{>{\centering\let\newline\\\arraybackslash\hspace{0pt}}m{#1}}
\newcolumntype{R}[1]{>{\raggedleft\let\newline\\\arraybackslash\hspace{0pt}}m{#1}}
\usepackage[table]{xcolor}
\usepackage{booktabs}
\usepackage{multirow}
\usepackage[colorlinks,citecolor=red,urlcolor=blue,bookmarks=false,hypertexnames=true]{hyperref} 
\usepackage{doi} 
\colorlet{tableheadcolor}{gray!25}
\colorlet{tablerowcolor}{gray!12.5}
\usepackage{tablefootnote}
\usepackage{siunitx}
\usepackage{lipsum}
\usepackage{subfig}
\usepackage[normalem]{ulem}

\usepackage[switch]{lineno}

\newcommand{\fp}{\ensuremath{F_{\text{prompt}}}}
\newcommand{\lr}{\ensuremath{L_{\text{recoil}}}}

\newcommand{\fpsc}{\ensuremath{F^{\text{nsc}}_{\text{prompt}}}}
\newcommand{\fpqpe}{\ensuremath{F^{\text{qpe}}_{\text{prompt}}}}

\newcommand{\lrsc}{\ensuremath{L^{\text{nsc}}_{\text{recoil}}}}
\newcommand{\lrqpe}{\ensuremath{L^{\text{qpe}}_{\text{recoil}}}}

\newcommand{\pleak}{\ensuremath{P_{\text{leak}}}}

\newcommand{\nsc}{\ensuremath{n_{\text{Sc}}}}
\newcommand{\nap}{\ensuremath{n_{\text{AP}}}}
\newcommand{\npe}{\ensuremath{n_{\text{PE}}}}
\newcommand{\qpe}{\ensuremath{\text{q}_\text{PE}}}

\newcommand{\totalpe}{\ensuremath{N_{\text{PE}}}}
\newcommand{\totalqpe}{\ensuremath{N_{\text{qpe}}}}
\newcommand{\totalnsc}{\ensuremath{N_{\text{nsc}}}}

\newcommand{\pter}{p(t)_{\text{er}}}
\newcommand{\ptnr}{p(t)_{\text{nr}}}

\newcommand{\Nuc}[2]{\ensuremath{^{#2}\mbox{#1}}}
\newcommand{\ar}{\Nuc{Ar}{39}}
\newcommand{\arr}{\Nuc{Ar}{40}}

\newcommand{\livedays}{840}
\newcommand{\lynsc}{6.05}
\newcommand{\lyqpe}{7.14}
\newcommand{\eventnumber}{$2.9\cdot 10^9$}

\graphicspath{{figures/}}

\journalname{Eur. Phys. J. C}
\title{Pulseshape discrimination against low-energy Ar-39 beta decays in liquid argon with 4.5 tonne-years of DEAP-3600 data}

\begin{document}
\author{P.~Adhikari~\thanksref{Carleton} 
        \and 
        R.~Ajaj~\thanksref{Carleton, Mcdonaldinst}
        \and
        M.~Alp\'{\i}zar-Venegas~\thanksref{UNAM}
        \and
        P.-A.~Amaudruz~\thanksref{Triumf}
        \and
        D.J.~Auty~\thanksref{Alberta, Mcdonaldinst}
        \and
        M.~Batygov~\thanksref{Laurentian}
        \and
        B.~Beltran~\thanksref{Alberta}
        \and
        H.~Benmansour~\thanksref{Queens}
        \and
        C.E.~Bina~\thanksref{Alberta, Mcdonaldinst} 
        \and 
        J.~Bonatt~\thanksref{Queens}
% ...
        (DEAP  Collaboration)\thanksref{email}
}

\thankstext{current}{Currently: Nuclear Science Division, Lawrence Berkeley National Laboratory, Berkeley, CA 94720}
\thankstext{current2}{Currently: Nikhef and the University of Amsterdam, Science Park, 1098XG Amsterdam, Netherlands}
\thankstext{email}{deap-papers@snolab.ca}
\thankstext{dec}{Deceased.}

\institute{Department  of  Physics,  University  of  Alberta,  Edmonton,  Alberta,  T6G  2R3,  Canada \label{Alberta}
% ...
\and 
    Department  of  Physics,  Carleton  University,  Ottawa,  Ontario,  K1S  5B6, Canada \label{Carleton}
% ...
\and 
    Department  of  Physics  and  Astronomy,  Laurentian  University,  Sudbury,  Ontario,  P3E  2C6,  Canada \label{Laurentian}
    \and
    Instituto de F\'{i}sica, Universidad Nacional Aut\'{o}noma de M\'{e}xico, A. P. 20-364, M\'{e}xico D. F. 01000, Mexico \label{UNAM}
% ...
\and
    Department of Physics, Engineering Physics and Astronomy, Queen's University, Kingston, Ontario, K7L 3N6, Canada \label{Queens}
% ...
\and
    TRIUMF, Vancouver, British Columbia, V6T 2A3, Canada \label{Triumf}
% ...
\and 
    Arthur B. McDonald Canadian Astroparticle Physics Research Institute, Queen's University, Kingston, ON, K7L 3N6, Canada \label{Mcdonaldinst}
% ...
}

\maketitle


\begin{abstract}
The DEAP-3600 detector searches for the scintillation signal from dark matter particles scattering on a 3.3 tonne liquid argon target. The largest background 
comes from \ar{} beta decays and is suppressed using pulseshape discrimination (PSD).

We use two types of PSD algorithm: the prompt-fraction, which considers the fraction of the scintillation signal in a narrow and a wide time window around the event peak, and the log-likelihood-ratio, which compares the observed photon arrival times to a signal and a background model. We furthermore use two algorithms to determine the number of photons detected at a given time: (1) simply dividing the charge of each PMT pulse by the charge of a single photoelectron, and (2) a likelihood analysis that considers the probability to detect a certain number of photons at a given time, based on a model for the scintillation pulseshape and for afterpulsing in the light detectors.

The prompt-fraction performs approximately as well as the log-likelihood-ratio PSD algorithm if the photon detection times are not biased by detector effects. We explain this result using a model for the information carried by scintillation photons as a function of the time when they are detected.

\end{abstract}


\section{Introduction} \label{sec:intro}

Liquid argon (LAr) is used as target material in several current and planned experiments related to the direct search for WIMP dark matter and studies of neutrino properties \cite{Fiorillo:2006bt,Gary:2013bj,Agostini2015jf,Meyers:2015dc,Aalseth:2017um,gerda,legend,Collaboration:2016ty,Akimov:2018ghi,Ajaj:2019wi}. Argon provides excellent pulse shape discrimination (PSD) power, which allows separation of backgrounds in the form of electron recoils\footnote{The term refers to events from particles that cause ionization and excitation mainly by interacting with the electrons in the argon atom, e.g. electrons, photons, or muons.} (ER) from the WIMP scattering signal expected to manifest as a nuclear-recoil (NR) event \cite{Fiorillo:2006bt,Agostini2015jf,Aalseth:2017um,Collaboration:2016ty,detectorpaper}.

 PSD in LAr is based on the large difference in lifetimes between the argon excimer's singlet (approximately \SI{6}{\nano\second}) and triplet (approximately \SI{1400}{\nano\second}) states. Due to this difference, the singlet-to-triplet ratio, which depends on linear energy transfer (LET) of the exciting particle \cite{Doke:1988uo}, can be determined very well on an event-by-event basis.

The dominant background in LAr-based dark matter detectors is the beta-decay of \ar, at a rate of approximately \SI{1}{\becquerel/\kilogram} in atmospheric argon \cite{Loosli:1983tl,Benetti:2007,Calvo:2018ho}. For a 3 tonne-year exposure, this amounts to $\mathcal{O}(10^8)$ events per keV in the region of interest for dark matter search. PSD is well-suited to suppress this background, as the dark matter-induced NR events have a different LET and thus different singlet-to-triplet ratio than the beta-induced ER events. 

Two types of methods are commonly used for PSD in liquid scintillators that show a double-exponential pulseshape: the prompt-fraction or tail-fraction, where the amount of light collected in a prompt or a late window is compared to the total amount collected, and a likelihood analysis where the photon detection times are compared to a signal and a background template. 
The implementation of both methods in the context of LAr scintillation detectors with $\mathcal{O}(10 $~kg) target mass has been discussed in \cite{Lippincott:2008uaa}, \cite{Akashi-Ronquest:2014jga}, and in \cite{d1psdpaper} we derived a first-principles model for the statistical distribution of the prompt-fraction PSD parameter. Here, we show the performance of both methods using \livedays~livedays of data from the 3.3~tonne target mass DEAP-3600 detector.

\section{The DEAP-3600 detector} \label{sec:experiment}
We give a brief overview of the DEAP-3600 detector in this section and refer the reader to \cite{detectorpaper} for a more detailed description.

The DEAP-3600 detector is located \SI{2}{km} underground at SNOLAB in Sudbury, Canada. The centre of the detector is a spherical volume \SI{170}{cm} in diameter, which contains \SI{3.3}{tonnes} of LAr. The scintillation light created in the LAr travels through the argon volume until it reaches the surface of the acrylic containment vessel. The inside acrylic surface is coated with TPB, which shifts the \SI{128}{\nano\meter} scintillation light to the blue spectral region. The wavelength-shifted light is transmitted to 255 Hamamatsu R5912 high quantum-efficiency PMTs through a total of \SI{50}{cm} of acrylic comprising the wall of the acrylic vessel and the acrylic light guides. The acrylic contains a UV-absorbing additive to suppress Cherenkov light produced in the vessel and light guides.

The inner detector is sealed inside a stainless steel sphere and suspended in an \SI{8}{m} diameter water shielding tank. The tank serves as active muon veto and as passive neutron moderator.

The signals from the 255~PMTs are passed to custom amplifiers which stretch and amplify them, before digitization by CAEN V1720 digitizers at \SI{4}{ns} resolution; enough to resolve single-photoelectron (SPE) pulses. Each PMT channel has a threshold of approximately \SI{0.1}{photoelectrons} (PE). To trigger the detector, a charge approximately equal to at least \SI{19}{PE} must be detected across all PMTs in a \SI{177}{ns} sliding window. Upon triggering, data from all PMTs are digitized from \SIrange{-2.6}{+14.4}{\micro\second} relative to the trigger time. The digitized traces are written to disk for offline analysis.

\section{Data selection and analysis}\label{sec:analysis}
\subsection{Data selection} \label{subs:dataselection}

The data used in this work represent approximately \livedays~livedays (4.5 tonne-years after event position cuts) of DEAP-3600 data taken between November~2016 and March~2020. The vast majority of recorded events are from \Nuc{Ar}{39} $\beta$-decays, with a small contribution of $\gamma$'s from radioactivity in detector materials \cite{EMBackground}. About 50\% of the livetime comes from data where events within and near the dark matter signal region have been removed from analysis for blinding purposes. The blinding removes a negligible amount of \Nuc{Ar}{39} events and will be discussed in more detail in Sect.~\ref{subs:systematics}. 



Events are selected using the following data cleaning cuts: 
\begin{enumerate}
    \item Low level DAQ cuts
    \item Timing: event peak must be near the DAQ trigger time
    \item Position: the reconstructed event position must be within the LAr volume and at least \SI{13}{cm} away from the inner detector surface
    \item Pile-up: at least \SI{20}{\micro\second} passed since the previous trigger, no more than three photons were detected in a time window from \SIrange{-2.6}{-1.6}{\micro\second} before the event peak, and no more than one event peak is recorded within the digitization time window
\end{enumerate}

These cuts do not remove all backgrounds. For example, a population of alpha decays with degraded energy is expected in the Dark-Matter signal region \cite{Ajaj:2019wi}. The full set of background cuts reduces the acceptance significantly, so to obtain the best possible statistics for ER events, we do not apply all cuts used in the WIMP search analysis.

A sample of WIMP recoil events is obtained through a Monte Carlo simulation (MC) of $\arr$ recoils distributed uniformly in the LAr with a flat energy spectrum. The simulation contains the full response of the detector including scintillation, photon scattering, wavelength shifting, PMT instrumental effects, and DAQ instrumental effects. The recoil quenching factor and the singlet/triplet ratios as a function of energy are taken from the SCENE measurements \cite{Cao:2015ue} at zero electric field. Validation of the WIMP simulation is described in \cite{Ajaj:2019wi}, and systematic uncertainties due to mis-modelling of $\arr$ recoil events are discussed in Sect.~\ref{subs:systematics}.

\subsection{Photon counting}

We use two methods to count the scintillation photons in a PMT pulse. 
A standard pulse-finding algorithm yields the peak time and the area under the peak for each pulse from the PMTs \cite{McElroy:2018vd}. A robust but naive way to count the number of photons in such a pulse is to divide the area under the peak, $Q$, by the mean charge of an SPE pulse, $Q_\text{SPE}$. We denote the number of PE obtained from charge division as \qpe. 
\begin{equation}
\qpe = \frac{Q}{Q_\text{SPE}}
\end{equation}

\qpe{} is a biased estimator for the number of scintillation photons. The bias is caused mainly by correlated noise in the PMT, in the form of afterpulsing (AP). The DEAP-3600 PMTs have a \qpe-weighted AP probability of approximately \SI{10}{\percent} \cite{pmtpaper,detectorpaper}. Since AP occurs in specific time windows between \SI{100}{\nano\second} and \SI{10000}{\nano\second}, it modifies the pulseshape \cite{DEAPCollaboration:2020hx}, and since AP is a statistical process, it also contributes to the variance of the \qpe{} count. The bulk of the variance in \qpe{}, however, is due to the width of the SPE charge distribution. For DEAP-3600 PMTs this width is approximately $\sigma_\text{SPE}/Q_\text{SPE} \approx 0.43$.

A less biased estimate of the number of scintillation photons can be obtained using a likelihood analysis. The method is described in \cite{Akashi-Ronquest:2014jga}, and the implementation in DEAP-3600 is explained in detail in \cite{Butcher:2016wy,McElroy:2018vd,Burghardt:2018tc}, and briefly summarized here. We assume that the number of PE in a pulse (\npe) is composed of PE from scintillation photons (\nsc) and signals from AP (\nap). The time response of the wavelength shifter and PMT dark noise are minor effects here and not considered. Using Bayes' theorem, we calculate how likely it is to observe \npe{} at the time of each pulse, given the pulse charge, the LAr scintillation probability density function (PDF), the times of preceding pulses, the AP time and charge PDF, and the SPE charge distribution of the PMT. For a pulse observed at a given time in the waveform, the posterior probability for \npe\ is 
\begin{equation}
P(\nsc + \nap=\npe|Q) = \frac{p(Q|\npe)p(\nsc)p(\nap)}{p(Q)}\;.\label{posterior}
\end{equation}
The values for $\nsc$ and $\nap$ are determined by finding the mean over Eq.~\eqref{posterior}\cite{Burghardt:2018tc}. The $\nsc$ values are real rather than natural numbers in this approach. The method differs slightly from what is described in \cite{Akashi-Ronquest:2014jga} and \cite{Butcher:2016wy}, where $\nsc$ and $\nap$ are estimated by taking the maximum of the posterior. When taking the maximum, there are regions in the pulseshape where a pulse is always more likely to originate from AP, and in these time regions, the algorithm will remove all scintillation photons.

\subsection{Energy and position reconstruction}

The event window in DEAP-3600 goes from \SIrange{-28}{+10000}{\nano\second} relative to the time of the event peak T$_0$. We denote the total number of PE in the event window as \totalpe, or if referring to a particular PE counting method, as \totalqpe{} or \totalnsc. T$_0$ and the event position in the detector are determined respectively based on the photon detection times across the PMT array and on the pattern of detected photons \cite{Ajaj:2019wi}. 

\totalpe{} is related to the energy deposited in the detector through the light yield. For events in the energy region of interest (ROI) for dark matter search, the light yield is approximately \SI{\lynsc}{\nsc/keV_{ee}} or \SI{\lyqpe}{\qpe/keV_{ee}}. The small dependence of light yield on event energy is not relevant for the data considered here.  All event energies are given in electron equivalent energy, denoted by \SI{}{keV_{ee}}.

We will use \totalnsc{} as the energy estimator throughout this report.

\section{PSD parameter definitions}\label{sec:pp}

The different PSD methods each define algorithms for calculating a PSD Parameter (PP) based on the detection times of scintillation photons for each event. SD power is based on the fact that different types of interactions produce distinguishable photon detection PDFs. We denote these PDFs $\ptnr$ and $\pter$ for nuclear recoils and electron recoils respectively. $\ptnr$ and $\pter$ have a dependence on energy that is not explicitly written out in the equations.  A detailed discussion of $\pter$ can be found in \cite{DEAPCollaboration:2020hx}.

\subsection{Prompt fraction}

The prompt fraction PP is commonly used in LAr-based detectors and is defined as
\begin{equation}
\fp = \frac{\sum_{t>t_\text{start}}^{t<t_\text{prompt}} n(t)}{\sum_{t>t_\text{start}}^{t<t_\text{total}} n(t)} \label{eq:fprompt}
\end{equation}
where $n$ can be either $n=\qpe{}$ or $n=\nsc$, and all times are relative to the event peak time T$_0$.

The time $t_\text{prompt}$ is chosen such that the difference between the cumulative distribution functions of $\ptnr$ and $\pter$, normalized to the standard deviation of the ER event distribution under a given time window,
\begin{equation}
    \Delta (t_\text{prompt}) = \frac{1}{\sigma_\text{er}(t_\text{prompt}) }\int_{t_\text{start}}^{t_\text{prompt}} (\ptnr - \pter) dt\;,
\end{equation}
is maximised. For LAr pulseshapes, this is the case for $t_\text{prompt}$ in the range of \SIrange{60}{80}{\nano\second}\footnote{This is assuming the time resolution is good enough that all of the singlet light is detected inside the prompt time window.}. 

In DEAP-3600 $t_\text{start}$=\SI{-28}{\nano\second}, the prompt window goes up to $t_\text{prompt}$=\SI{60}{\nano\second}, and the full event window ends at $t_\text{total}$=\SI{10000}{\nano\second}. The denominator of Eq~\eqref{eq:fprompt} is thus equal to \totalpe. $t_\text{start}$ is negative so that photons that are detected before the event time due to the finite timing resolution of the detector are also counted.

This method gives us two PPs:
\begin{itemize}
    \item \fp{} based on \qpe{} (\fpqpe )
    \item \fp{} based on \nsc (\fpsc )
\end{itemize}

\fpqpe with a prompt time window of \SI{150}{\nano\second} was used in \cite{d1psdpaper} and in \cite{Amaudruz:2018gr}, while \cite{Ajaj:2019wi} used \fpsc{} with a \SI{60}{\nano\second} prompt window and \cite{darkside50_232d} used \fpqpe with a \SI{90}{\nano\second} prompt window.

\subsection{Log-likelihood-ratio}

The construction of the log-likelihood-ratio $\lr$ PP is described in \cite{Akashi-Ronquest:2014jga} and \cite{Burghardt:2018tc}.
Namely 
\begin{equation}
\lr = \frac{1}{2} + \frac{\sum_{t>t_\text{start}}^{t<t_\text{total}} w(t) n(t)}{\sum_{t>t_\text{start}}^{t<t_\text{total}} n(t)}
\label{eq:lrecoildef}
\end{equation}
with the weights defined as
\begin{equation}
w(t) = \frac{1}{2}\cdot \log \frac{\ptnr}{\pter}. \label{eq:lrecoilweights}
\end{equation}
The scaling and addition of a factor of $1/2$ in Eqs.~\eqref{eq:lrecoildef} and~\eqref{eq:lrecoilweights}  are chosen to force the values \lr{} can take to fall between zero and one.

We note that the Gatti parameter \cite{Gatti:BPrNIqM1}, which is also commonly used for PSD in scintillation detectors, is formally the same as the \lr{} parameter up to third order in $\frac{\ptnr}{\pter}$ \cite{Burghardt:2018tc}. It therefore yields similar PSD performance. 

The photon detection time PDFs $\ptnr$ and $\pter$ are created by first mixing the singlet and triplet component of the pulseshape with weights from \cite{Cao:2015ue} for nuclear recoils and weights as measured in DEAP-3600 for electron recoils. This step is necessary to account for the energy dependence of the singlet fraction. Next, the model PDF is convolved with the detector time resolution from \cite{pmtpaper} and the TPB fluorescence PDF from \cite{2015PhRvC..91c5503S}\footnote{For the shape of the slow TPB fluorescence, \cite{2015PhRvC..91c5503S} is superseded by \cite{Stanford:2018un}, but both shapes are very similar at the time scales of interest here.}. Finally, a flat dark noise component is added. This method can work either with \qpe{} or with \nsc. When using \qpe{}, $\pter$ and $\ptnr$ are also convolved with the AP PDF.

This method gives us two more PPs:
\begin{itemize}
    \item \lr{} based on \qpe{} (\lrqpe )
    \item \lr{} based on \nsc{} (\lrsc )
\end{itemize}

\section{Performance of the 4 PSD parameters}\label{sec:performance}

\subsection{The DEAP-3600 data in 4 PSD parameters}

\begin{figure}[htbp]
\centering
  \centering
  \includegraphics[width=\linewidth]{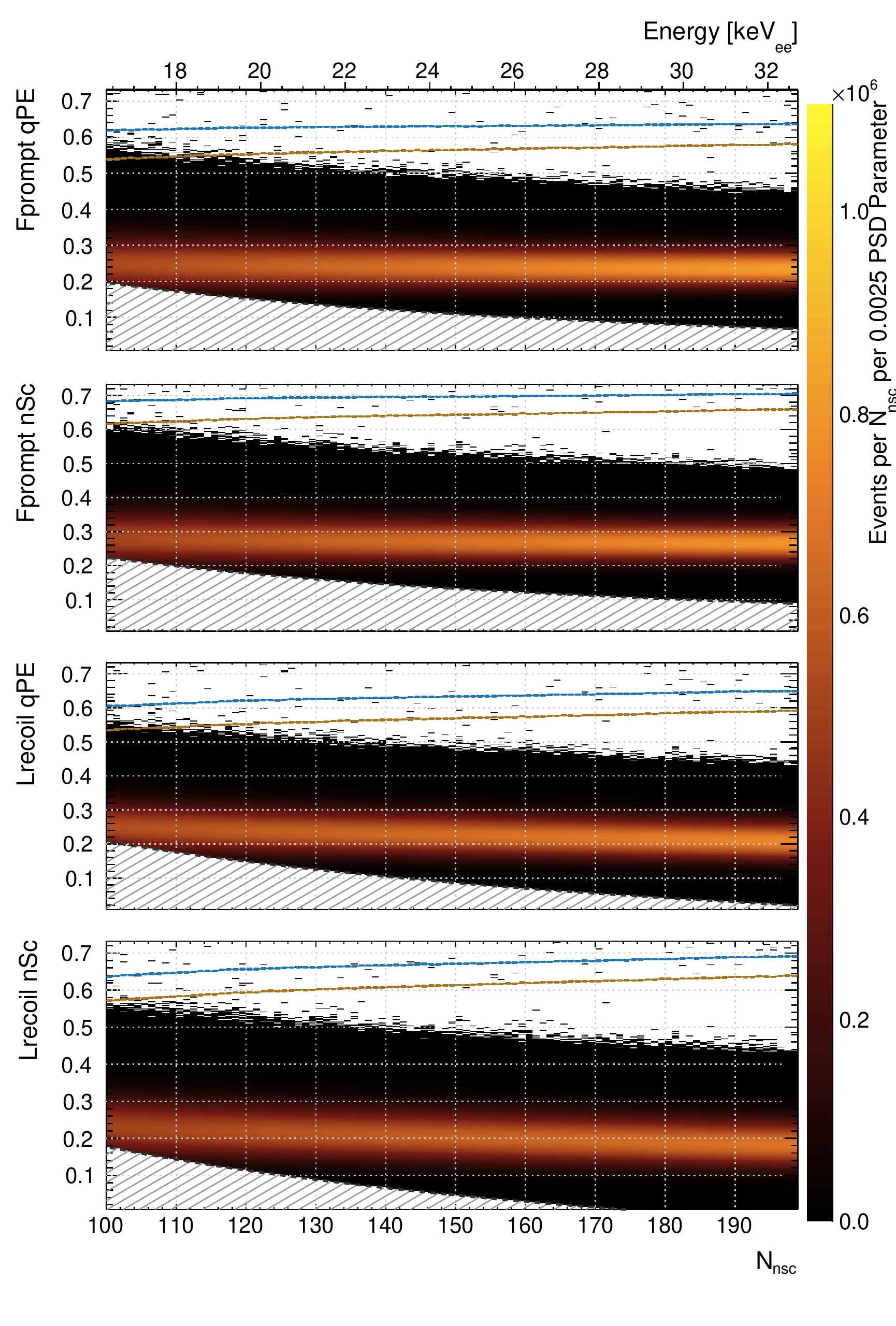}
  \caption{The distributions of mainly \ar{} $\beta$ decay events for each of the four PSD parameters as a function of energy. The 50\% (blue) and 90\% (brown) nuclear recoil acceptance lines are also shown. Some events are expected above the \ar{} population because only a subset of the WIMP analysis cuts are used.}
  \label{fig:summary2d}  
\end{figure}

Fig.~\ref{fig:summary2d} shows DEAP-3600 data between approximately \SIrange{100}{200}{\totalnsc}, that is \SIrange{16}{33}{keV_{ee}}, binned by \totalnsc{} on the x-axis and the four different PPs discussed in Sect.~\ref{sec:pp} on the y-axes.

The blue (brown) lines on each sub-figure are the values of the PP above which 50\% (90\%) of the nuclear recoil signal can be found, based on MC.

In the hatched regions, the trigger efficiency is less than 99.5\%. DEAP-3600 uses a threshold trigger on the total charge recorded in a sliding window of length \SI{177}{\nano\second} (see Sect.~\ref{sec:experiment}). In the variables available to the trigger, that is \qpe{} and \fpqpe{} with a prompt window of \SIrange{-28}{150}{\nano\second}, the trigger removes all events below the line described by
\begin{equation}
    \text{TE}(\totalpe) = \frac{N_{qpe}^\text{trigger}}{\totalpe} \label{eq:trigger}
\end{equation}
where $N_{qpe}^\text{trigger}$ is approximately \SI{19}{\qpe}. For offline variables like \totalnsc{} and all the PPs discussed here, the shape is more complicated. Furthermore, the details of the DEAP-3600 data acquisition system lead to a run- and time-dependent variation in the trigger efficiency, which is calibrated using the method described in \cite{Pollmann:2019fb} and inherently has a large uncertainty. The trigger efficiency is then corrected for with the run-specific calibration, event-by-event. By comparing the corrected to the uncorrected sum histograms over all data, the trigger efficiency in \totalnsc{} versus PP space on average across the data set is found.

Because the trigger efficiency has to be calibrated on a run-by-run basis, and a single run has many empty bins in regions of \totalnsc{} and \fp{} or \lr{} far from the peak, only regions near the peak of the PP distribution can be corrected. Due to these complications, the corrected histograms are only used to judge where the data are reliable, but do not replace the uncorrected data.

\subsection{The PP distribution model and fits}

We want to quantify how close the ER background population is to the NR signal population in Fig.~\ref{fig:summary2d} for each PP and as a function of \totalnsc. We denote the two distributions P$^{\text{ER}}(x, \totalnsc)$ and P$^{\text{NR}}(x, \totalnsc)$.

For a background-free WIMP search, a cut in the value of the PP must be chosen such that the probability for a background event to reconstruct as a signal event is $\ll 1$. In order to extrapolate the P$^{\text{ER}}(x, \totalnsc)$ distribution to all possible values of the PP, a 2-dimensional empirical model is fit to it. The model describes the PP distribution as a gamma distribution,
\begin{equation}
\Gamma(x; \mu, b) = \frac{1}{b\mu\Gamma\left(\frac{1}{b}\right)}\left(\frac{x}{b\mu}\right)^{\frac{1}{b}-1}e^{-\frac{x}{b\mu}} \label{eq:gamma}
\end{equation}
convolved with a Gaussian:
\begin{align}
\Phi(x,\totalnsc) = &I(\totalnsc) \cdot \Gamma (x ;\mu(\totalnsc), b(\totalnsc)) \\
   &\ast \text{Gauss}(x;\mu=0,\sigma(\totalnsc)).
\end{align}
For \fp{}, the energy dependence is introduced by making the shape parameters, $\mu$, $b$, and $\sigma$ functions of \totalnsc :
\begin{align}
b(\totalnsc) &= a_0 + \frac{a_1}{\totalnsc} + \frac{a_2}{\totalnsc^2} \\
\sigma(\totalnsc) &= a_3 + \frac {a_4}{\totalnsc} + \frac{a_5}{\totalnsc^2}  \\
\mu(\totalnsc)&= a_6 + \frac{a_7}{\totalnsc} + \frac{a_8}{\totalnsc^2} + \frac{a_9}{\totalnsc^3} 
\end{align}
 The functional form is empirical, and the parameters $a_0,...,a_{9}$ have no physical meaning. The normalization $I(\totalnsc)$ is not floated in the fit. In the evaluation of $\Phi(x,\totalnsc)$, $I(\totalnsc)$ is automatically chosen so that the function value matches the height of the data at the peak of the histogram. 

Turning the one-dimensional distribution into a two-dimensional one by requiring the shape parameters to evolve smoothly with \totalnsc{} is not strictly necessary. For the \lr{} parameters, each \totalnsc -slice is fit separately with the 1D model, because the behaviour of the shape parameters with \totalnsc{} is slightly more complicated.
However, performing individual 1D fits introduces statistical variation that leads to leakage predictions which are not perfectly smooth as a function of \totalnsc , which is undesirable in the context of constructing a region of interest for a WIMP search.

\begin{figure*}[htbp]
\centering
\subfloat[]{
\centering
  \includegraphics[width=\columnwidth]{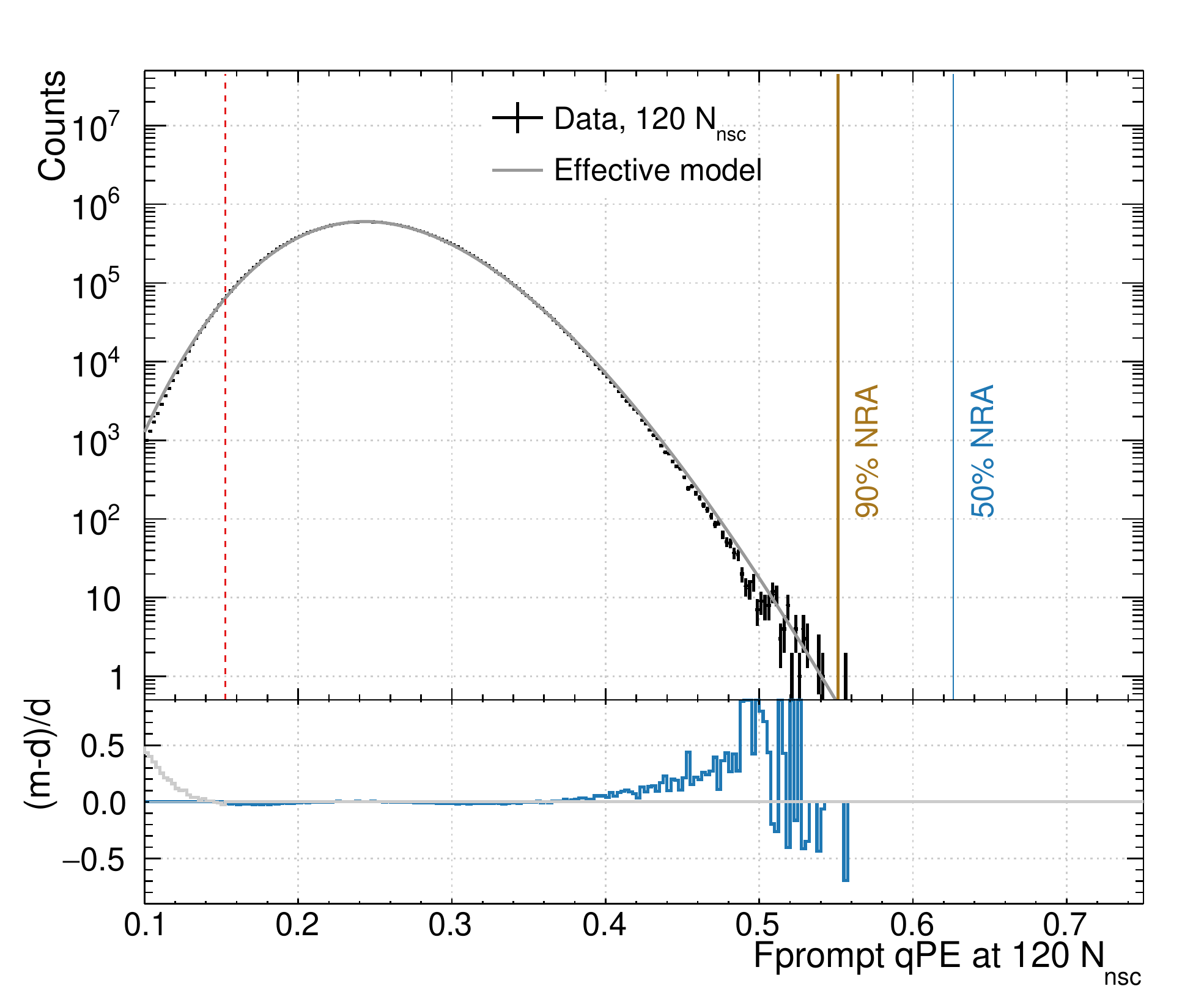}
}%
\subfloat[]{
\centering
  \includegraphics[width=\columnwidth]{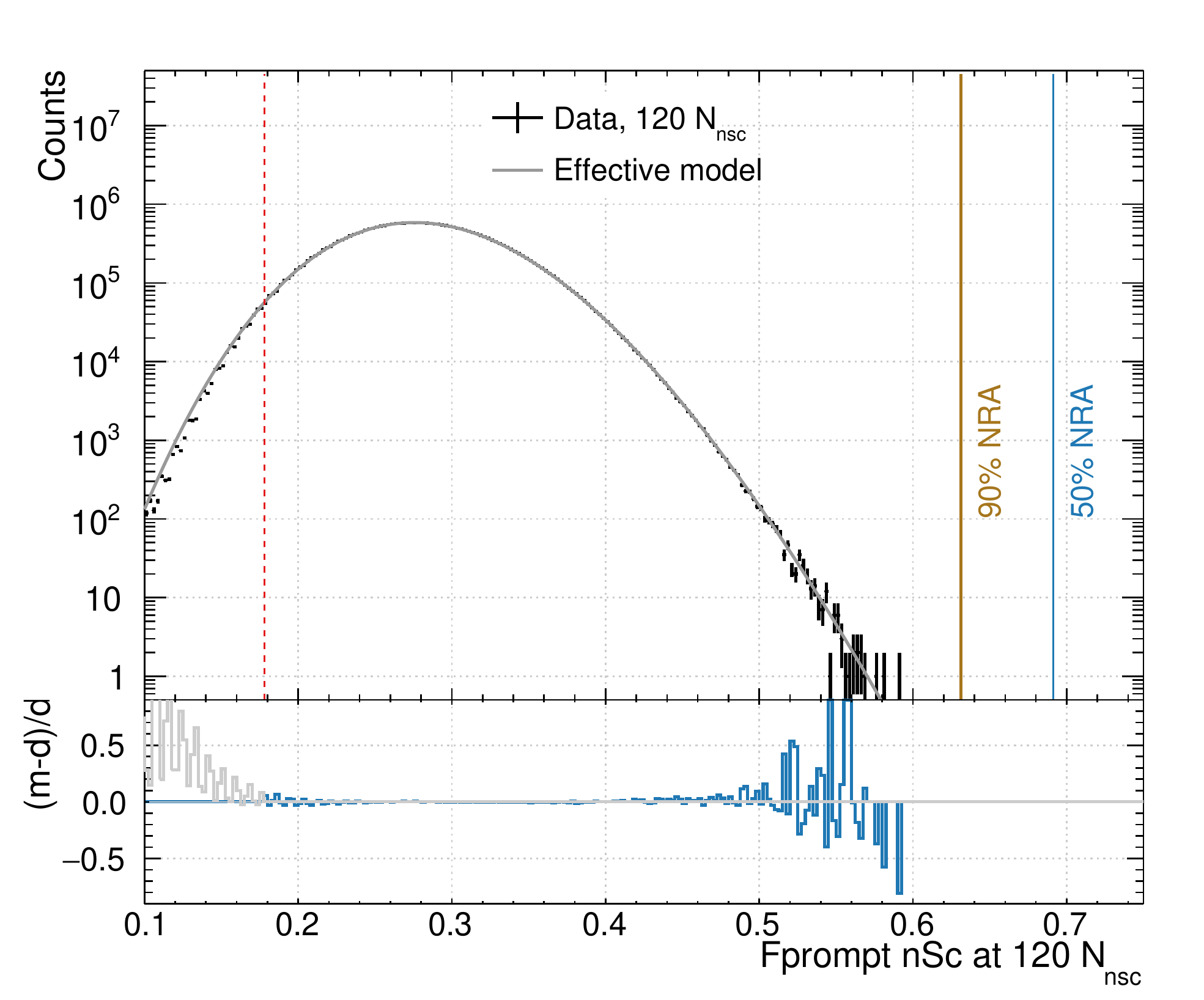}
}
\newline
\subfloat[]{
\centering
  \includegraphics[width=\columnwidth]{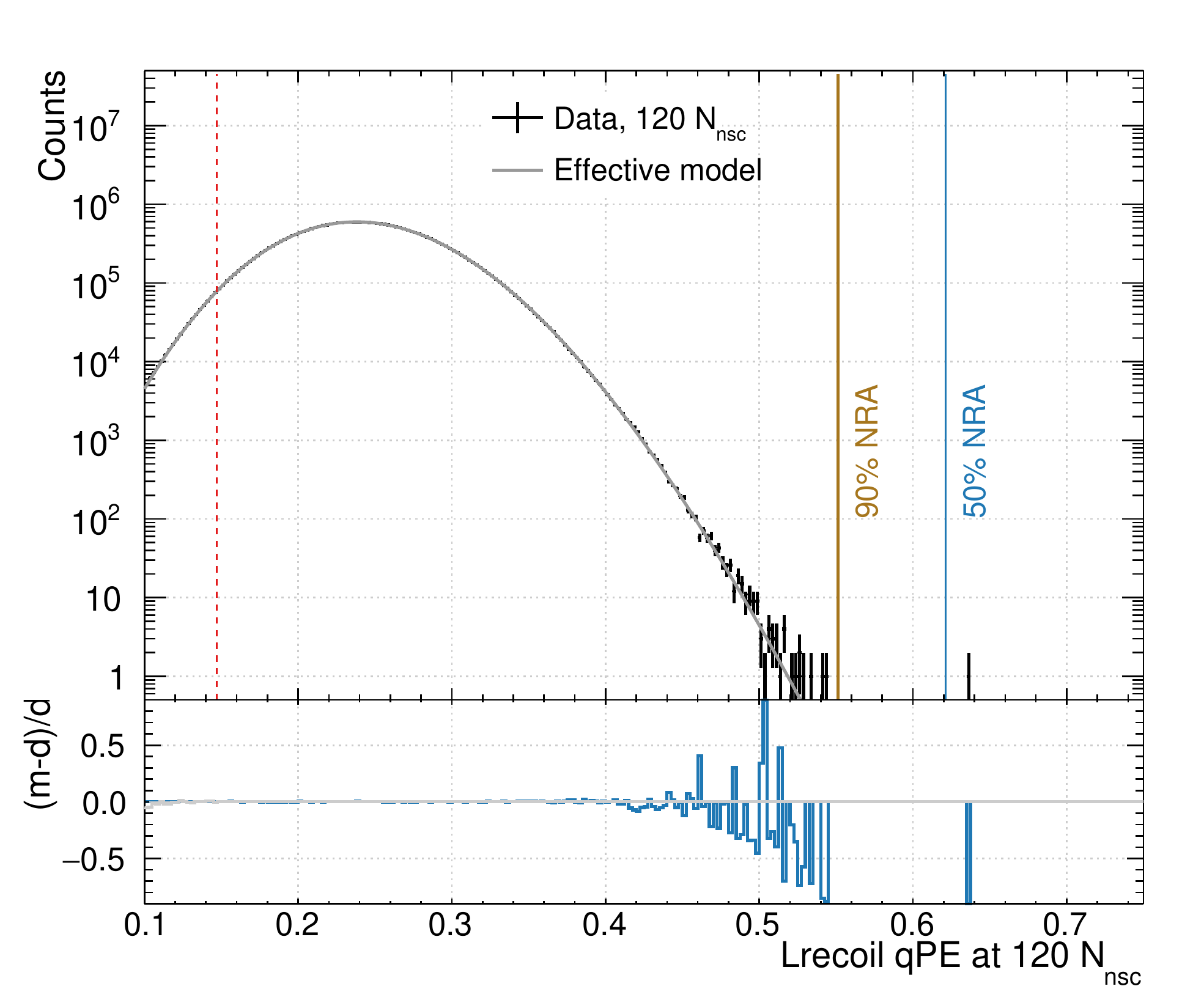}
}
\subfloat[]{
\centering
  \includegraphics[width=\columnwidth]{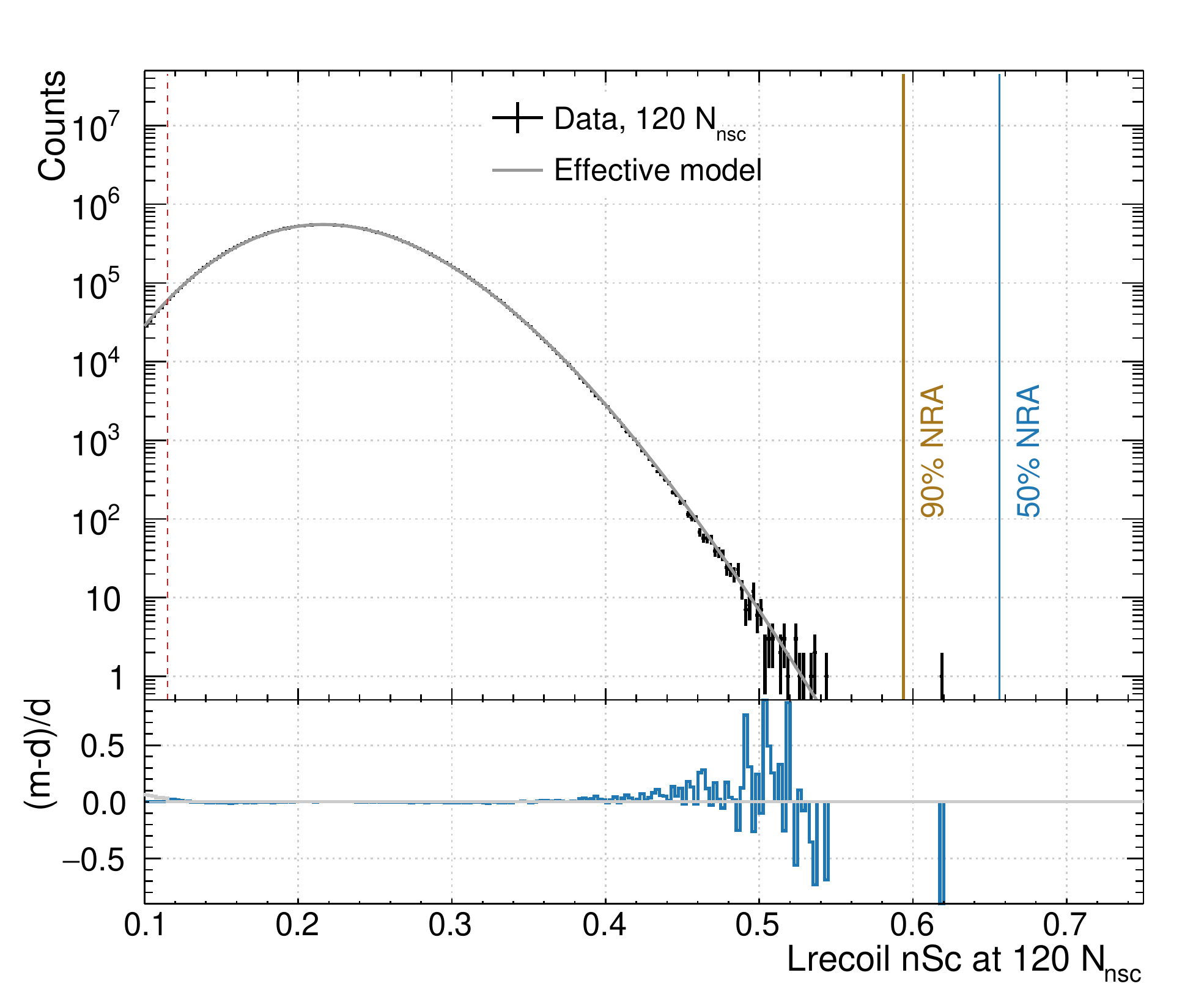}
}
\caption{The PP distributions for events at \SI{120}{\totalnsc} (approximately \SIrange{19.65}{19.82}{\kilo\electronvolt_{ee}}) are shown together with the effective model fits. The bin width is 0.0025 and there are $2.8\cdot10^7$ events in each histogram. The vertical dashed red line marks the \fp{} value above which the trigger efficiency is 99.5\%. The brown (blue) line marks a nuclear recoil acceptance of 90\% (50\%). The lower panel shows the relative deviation between the model and the data.}
\label{fig:Rprompt1D}
\end{figure*}

\begin{figure}[htbp]
  \centering
  \includegraphics[width=\columnwidth]{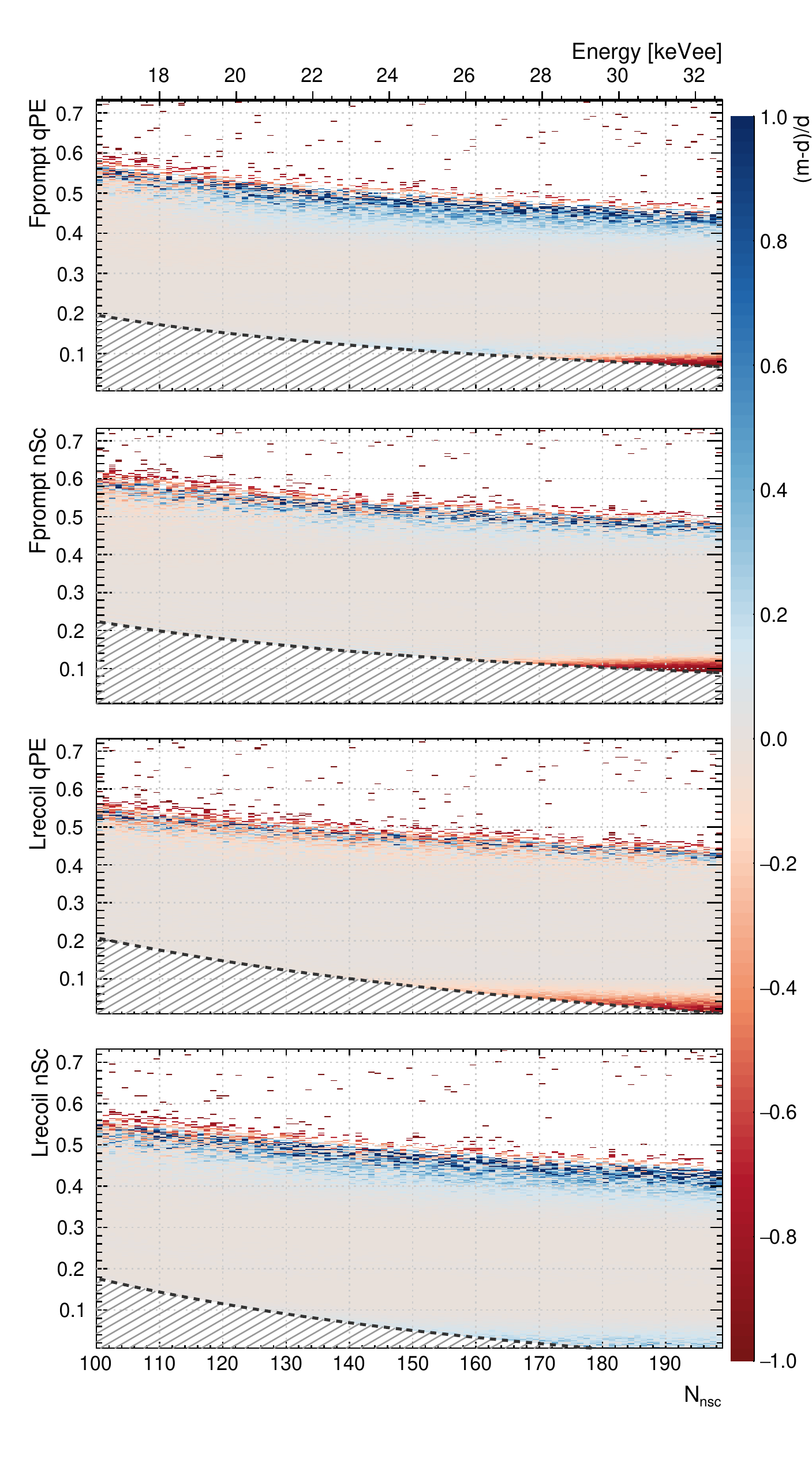}
  \caption{The relative difference, (model-data)/data, for each of the four PSD parameters. The only significant deviations between model and data are at the edge of the distributions, where several systematic effects bias the event count (see Sect.~\ref{subs:systematics}).}
    \label{fig:residuals2d}
\end{figure}

Fig.~\ref{fig:Rprompt1D} shows the fit results in a 1D slice at 120~\totalnsc (this includes events with a value of \totalnsc{} in [120, 121)). Fig.~\ref{fig:residuals2d} shows the relative difference between model and data over the 2D space for each PP. 
The fit is performed over a range on the energy axis of [98:210]~\totalnsc, and on the PP axis from the upper edge of the trigger-efficiency curve to 0.6.

\subsection{Leakage predictions}
We judge the performance of a PP by the probability of a background event to leak into the signal region as a function of \totalnsc{} and nuclear recoil acceptance (NRA). 
Fig.~\ref{fig:summary2d} shows the means of the background and signal distributions becoming closer toward lower energies, which is expected from the energy-dependence of the singlet to triplet ratio. The separation between the means of the populations is therefore a function of energy. The spread of the distributions about the mean is largely determined by counting statistics, that is by \totalnsc. Hence, it is natural to consider the leakage probability in slices that are one~\totalnsc{} wide.

The NRA is the fraction of signal events in the range $[x, 1]$,
\begin{equation}
    \text{NRA}(x, \totalnsc) = \frac{\int_{x}^1 P^{\text{NR}}(x', \totalnsc) dx'}{\int_{0}^1 P^{\text{NR}}(x', \totalnsc) dx'} \; .
\end{equation}

The uncertainty on the NRA is calculated using the binomial confidence interval (the Wilson score) at 95\% coverage.  

The fraction of ER background events that leaks into the region $[x, 1]$ is defined similarly as
\begin{equation}
\pleak(x,\totalnsc) = \frac{ \int_x^1 \Phi(x', \totalnsc)dx' }{\int_0^1 \Phi(x', \totalnsc)dx' }
\end{equation}
where we use the mathematical model $\Phi(x,\totalnsc)$ rather than the data $P^{\text{ER}}(x, \totalnsc)$.

The uncertainty on $\pleak(x,\totalnsc)$ is determined by assuming the parameter uncertainties returned by the fitter are Gaussian and randomly drawing parameter combinations from a multidimensional Gaussian with mean and standard deviation for each parameter and correlations between parameters taken from the fit result. For each parameter combination, $\pleak^i(x,\totalnsc)$ is calculated, where $i$ stands for the $i$\textsuperscript{th} set of randomly drawn parameters. For each x, the central value of the leakage probability is the mean over all $\pleak^i(x,\totalnsc)$ (this reproduces the curve obtained from the parameter values returned by the fit), and the up (down) uncertainties are the standard deviations among all cases where $\pleak^i(x,\totalnsc)$ is bigger (smaller) than the nominal value.

\begin{figure}[htbp]
\centering
\subfloat[]{
\centering
  \includegraphics[width=\columnwidth]{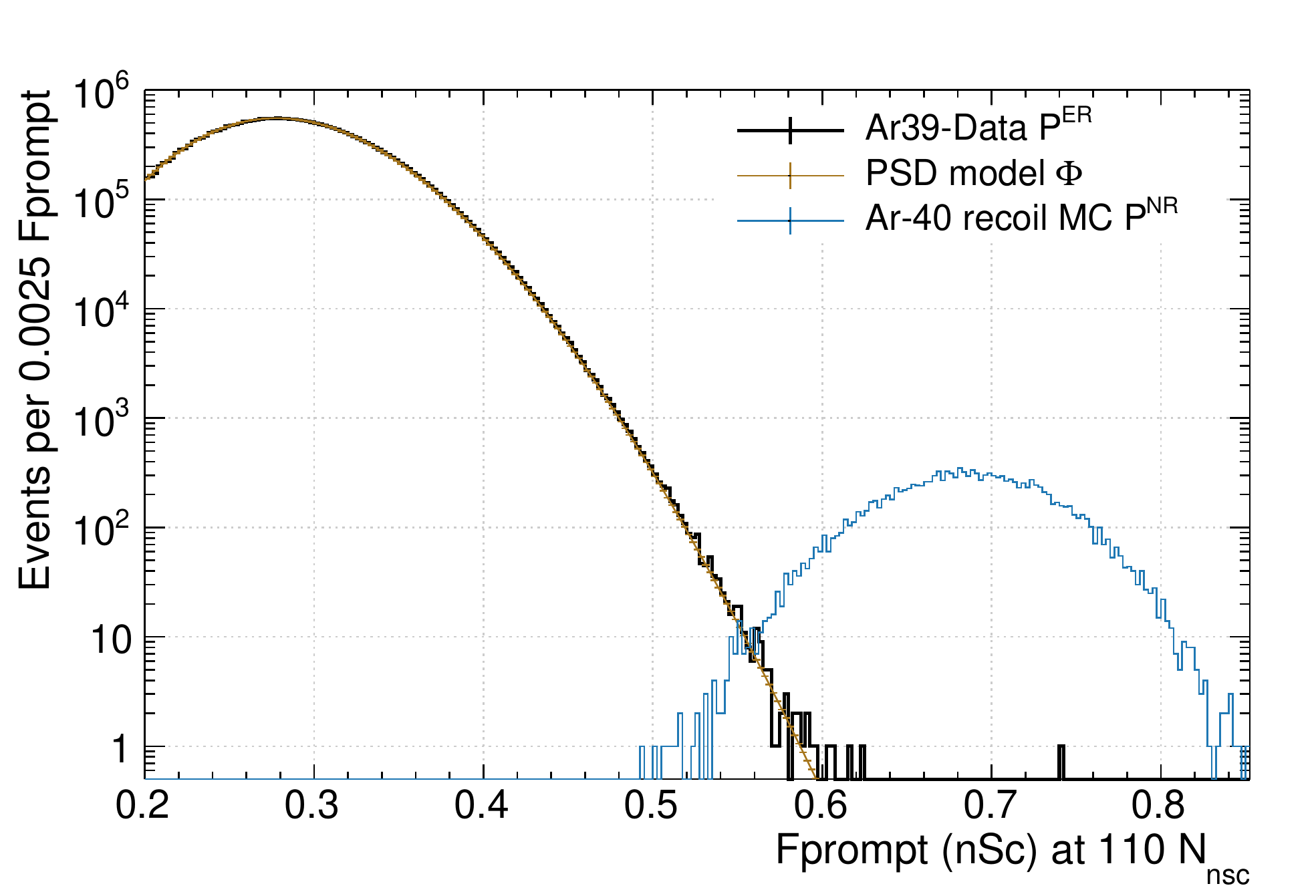}
} 
\newline
\subfloat[]{
\centering
  \includegraphics[width=\columnwidth]{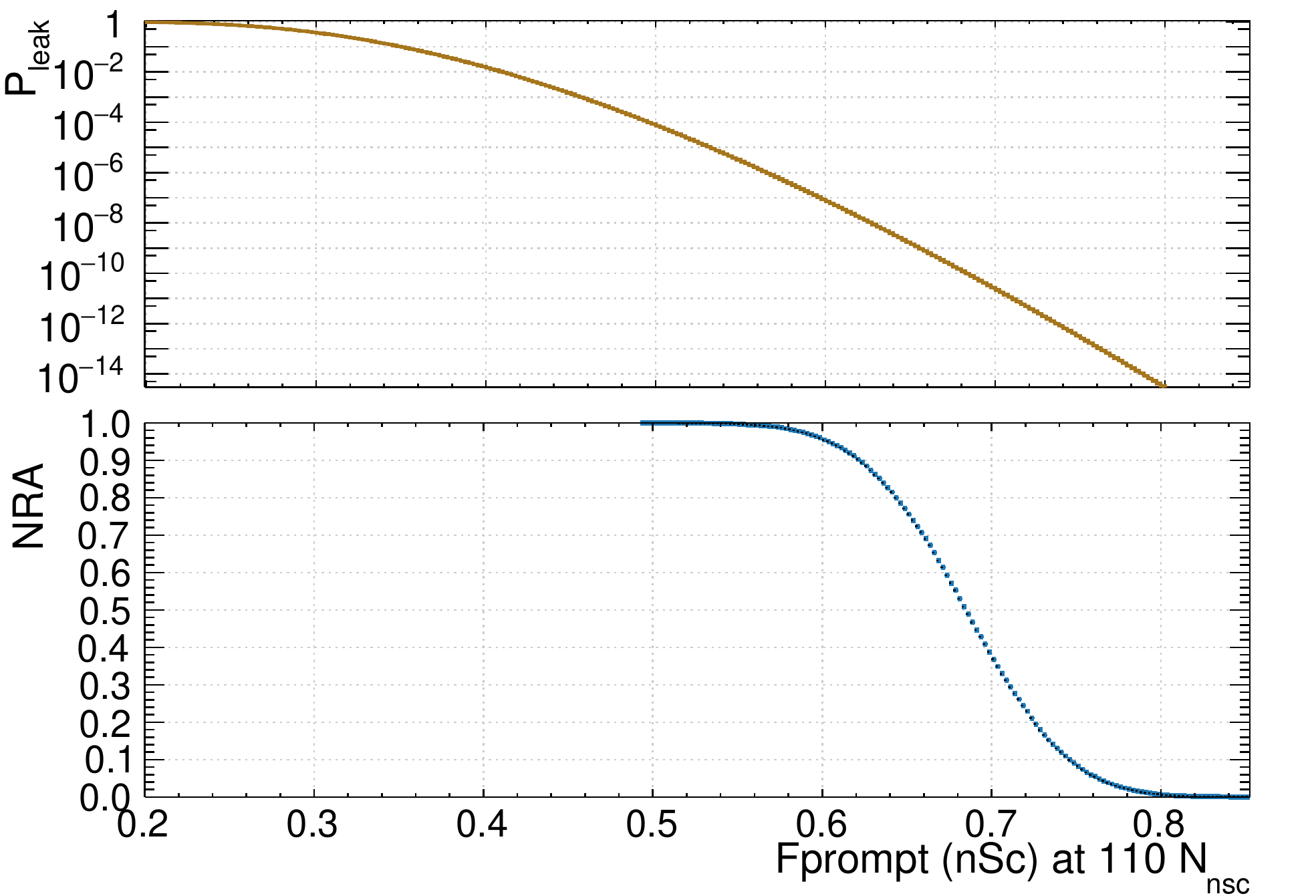}
}
\caption{(a) The \fpsc{} distributions at \SI{110}{\totalnsc} are shown for \Nuc{Ar}{39} $\beta$ events (background), together with the model fit, and for simulated \Nuc{Ar}{40} recoil events (signal). (b) The background leakage probability (based on the fit model to \ar{} data) and signal acceptance (based on signal MC) as a function of the PSD parameter is shown.}
\label{fig:psd_fprompt}
\end{figure}

Fig.~\ref{fig:psd_fprompt} (a) shows $P^{\text{ER}}$, $\Phi$, and $P^{\text{NR}}$, and (b) shows the respective \text{NRA} and $\pleak$, all for \fpsc{} and at \totalnsc=110.

\begin{figure}[htbp]
\centering
\subfloat[]{
\centering
  \includegraphics[width=\columnwidth]{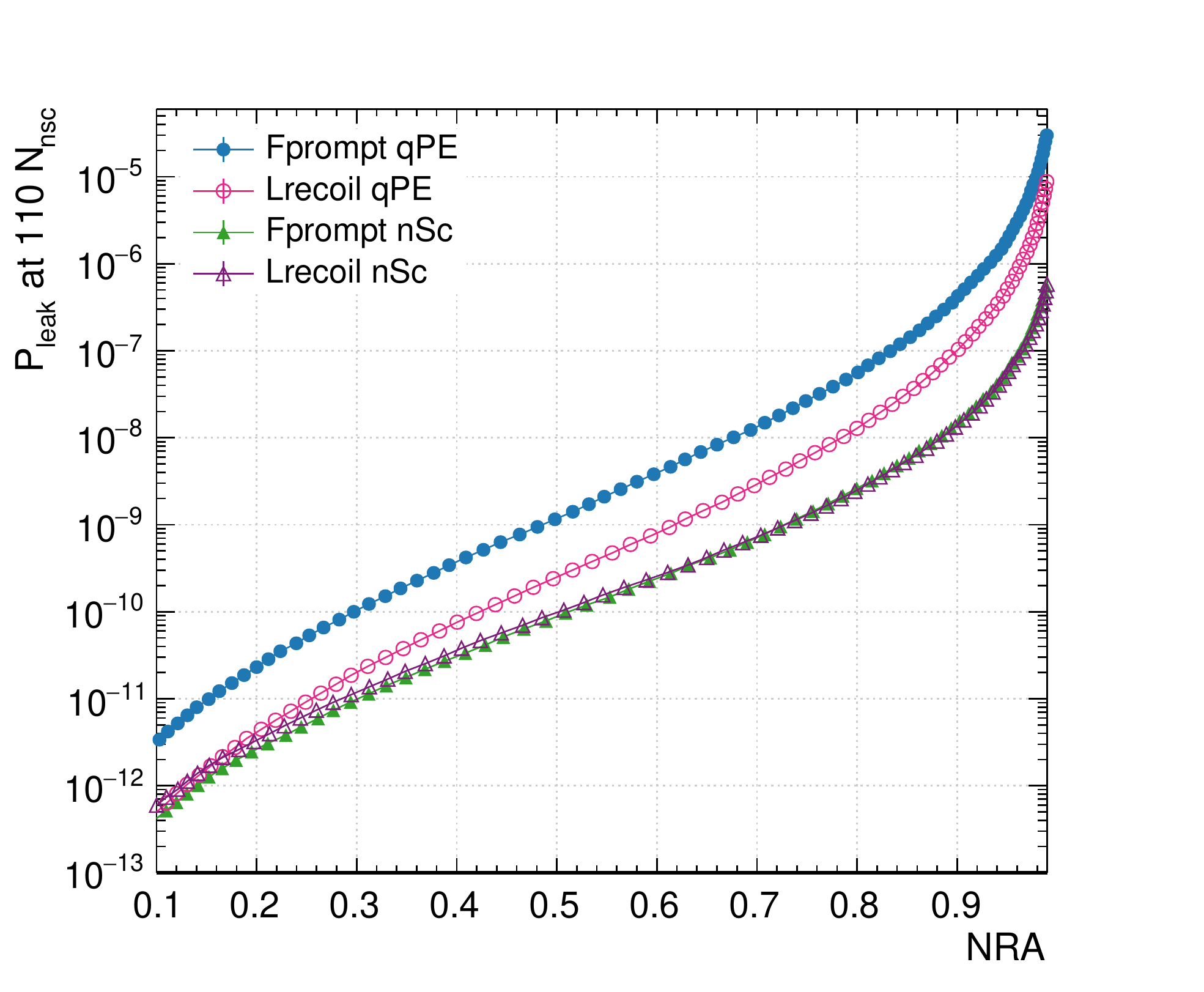}
}%
\newline
\subfloat[]{
\centering
  \includegraphics[width=\columnwidth]{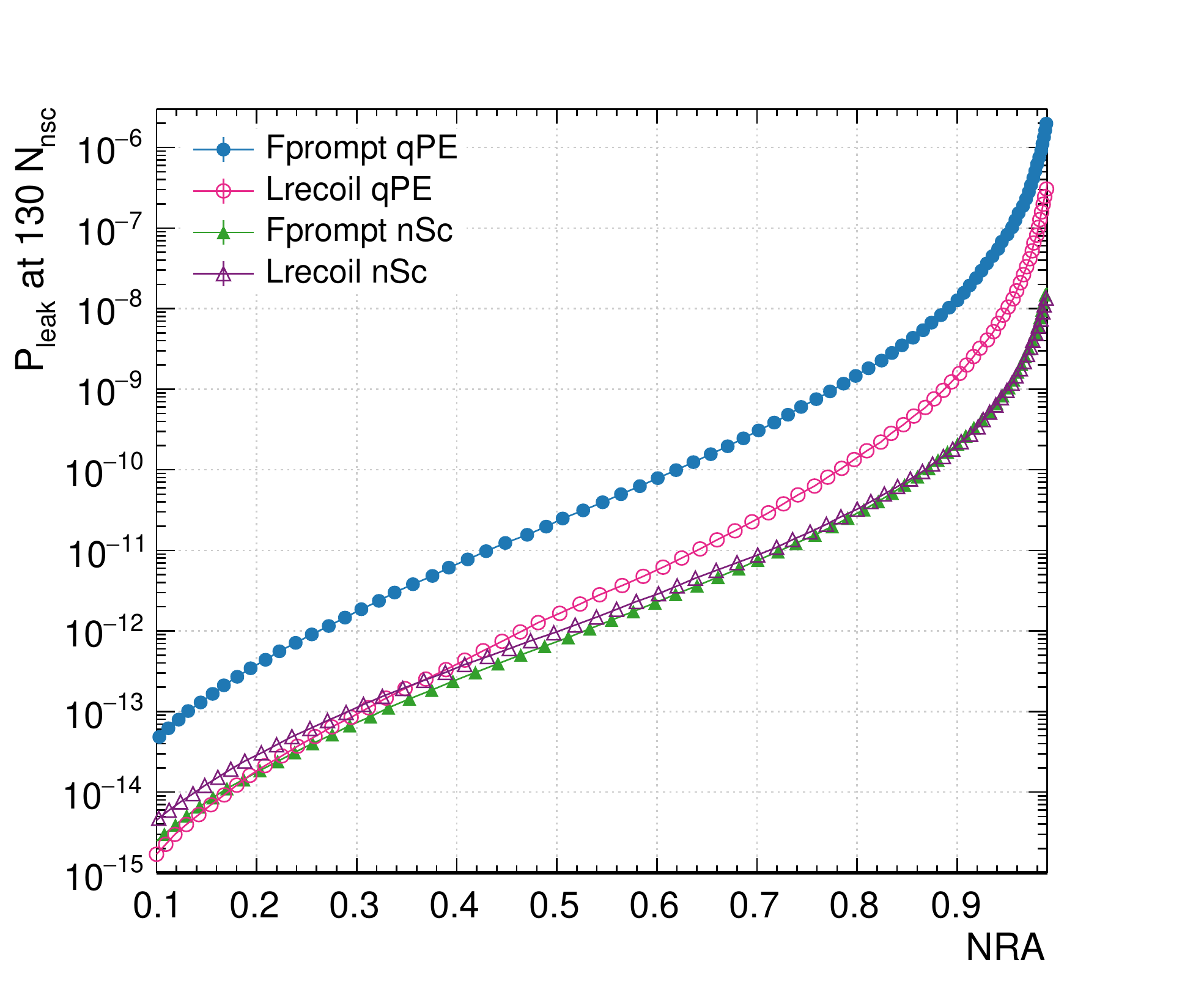}
}
\caption{Leakage probabilities for each PP as a function of NRA for events at (a) \SI{110}{\totalnsc} (approximately \SIrange{17.46}{17.61}{keV_{ee}}) and at (b) \SI{120}{\totalnsc} (approximately \SIrange{19.65}{19.82}{\kilo\electronvolt_{ee}}). Statistical error bars, where not visible, are smaller than the marker size.}
\label{fig:leakagevse}
\end{figure}

In Fig.~\ref{fig:leakagevse}, the leakage probability is shown as a function of NRA for all four PPs and at two values of \totalnsc. \SI{110}{\totalnsc} is close to the lower threshold relevant to the DEAP-3600 detector, and above \SI{130}{\totalnsc}, the separation between background and signal populations is so good that PSD leakage is no longer as relevant.

\begin{figure}[htbp]
\centering
\subfloat[]{
\centering
  \includegraphics[width=\columnwidth]{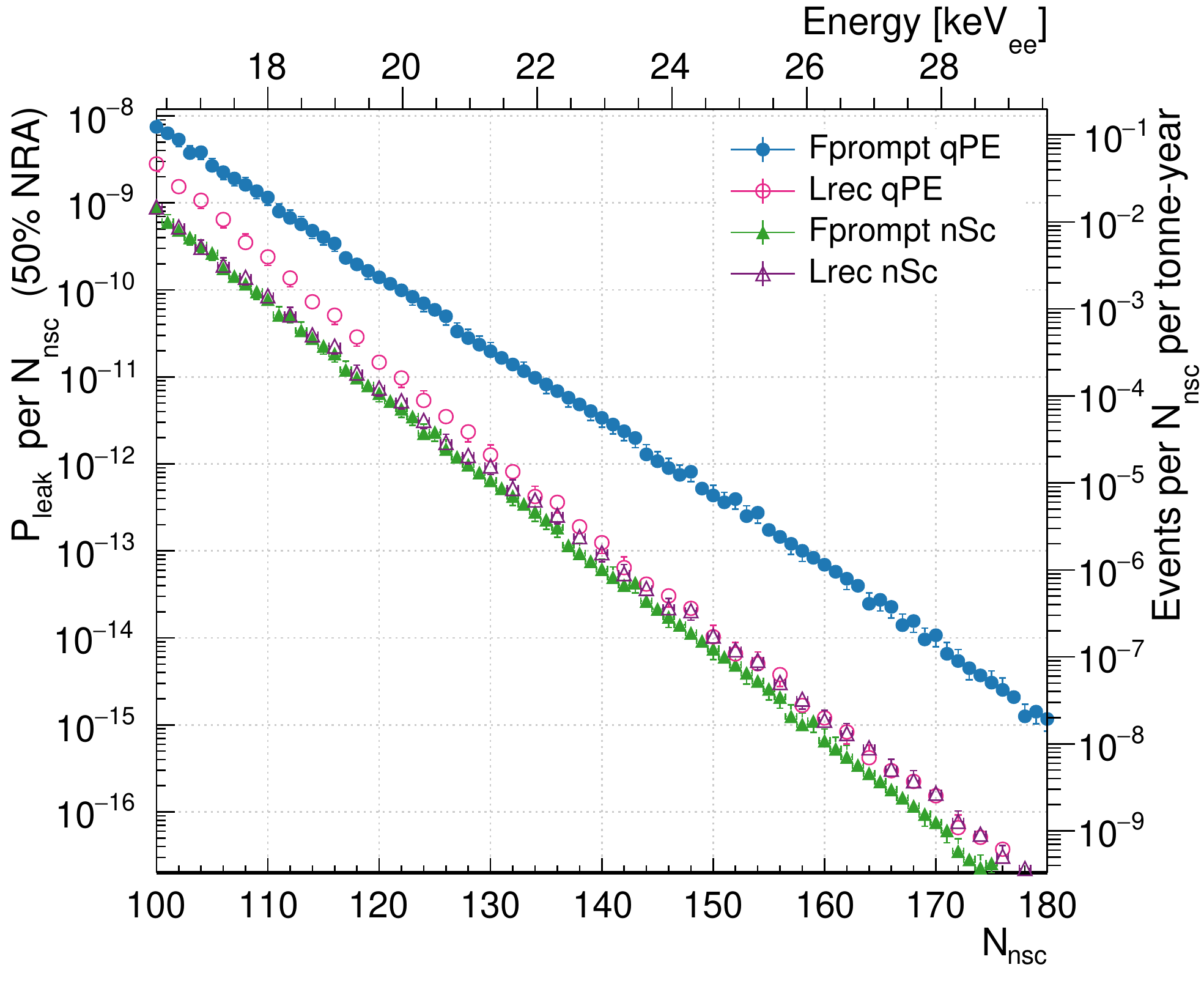}
}
\newline
\subfloat[]{
\centering
  \includegraphics[width=\columnwidth]{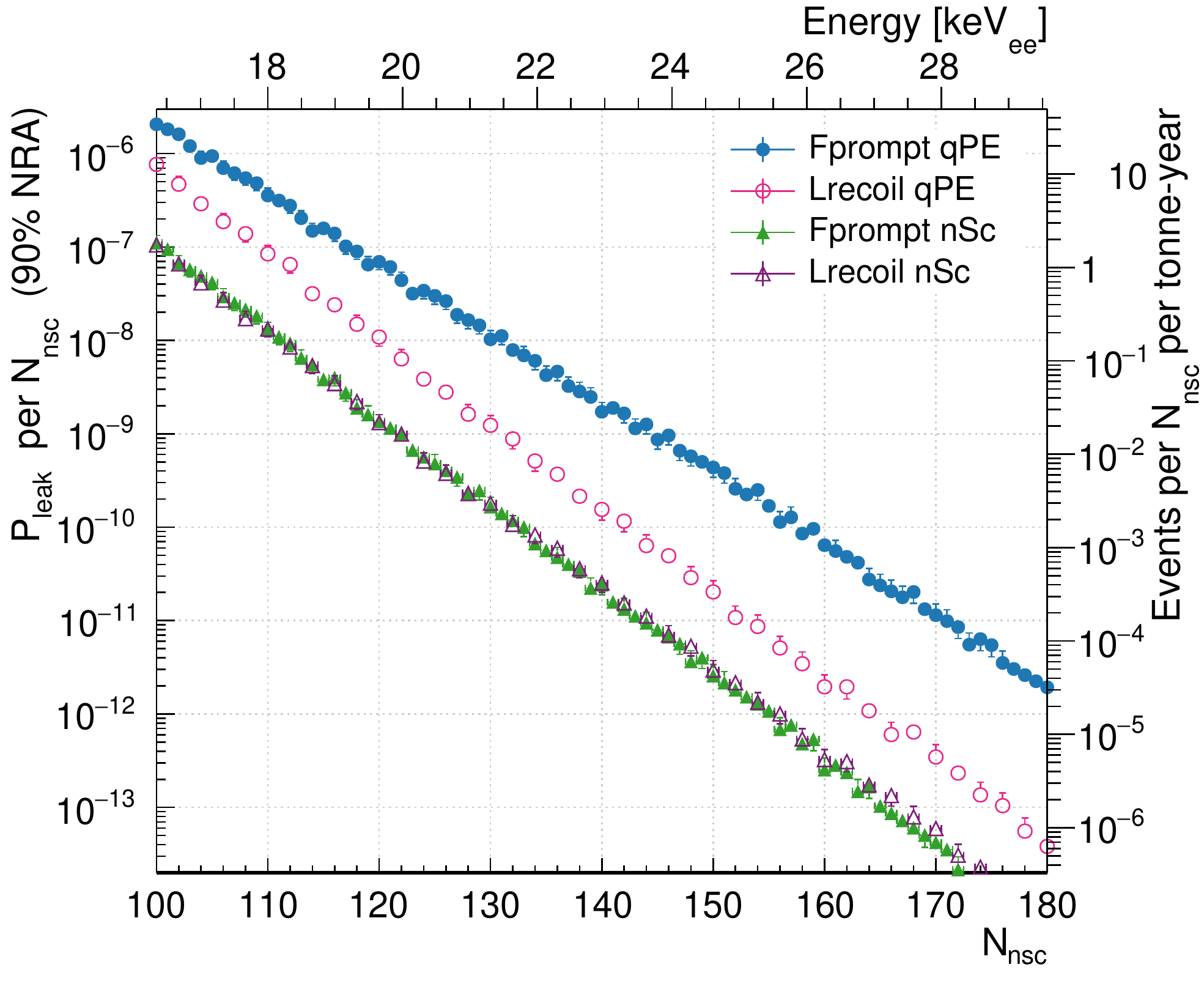}
}
\caption{The leakage probability as a function of \totalnsc{} at (a) 50\% and (b) 90\% NRA. Statistical error bars, where not visible, are smaller than the marker size.}
\label{fig:psd_withenergy}
\end{figure}

For 50\% and 90\% NRA, Fig.~\ref{fig:psd_withenergy} shows as a function of energy the leakage probability and the number of leaked events per \totalnsc{} this would correspond to in a nominal 1~tonne~year exposure. The error bars are dominated by the uncertainty in the NRA position and are correlated between bins. The statistical uncertainty from the fit is negligible.

\subsection{Systematic uncertainties} \label{subs:systematics}

\subsubsection{Pile-up}

The data contain pile-up events from a) random coincidences between two \ar{} decays b) random coincidences of \ar{} with Cherenkov light produced in the acrylic, and c) correlated pile-up between Cherenkov light in the acrylic or PMT glass and an ER event (other than \ar{} beta decay) in the LAr, for example from a gamma emitted by \Nuc{Tl}{208} in the PMT glass that interacts both in the acrylic and in the LAr. While LAr scintillation timing has a double-exponential decay time structure, Cherenkov light is emitted in a flash shorter than the time resolution of the detector.

\begin{table}[htp]
\caption{Estimated fractions of events between \SIrange{100}{200}{\totalnsc} that are piled-up with a second event and not removed by the cuts, for different types of coincidences. `Timing' distinguishes between coincidences that occur within the prompt \fp{} window and those that occur within the late \fp{} window.}
\begin{center}
\begin{tabular}{lll}
\hline
Partners & Timing & Fraction \\
 \hline
\multicolumn{3}{l}{Random coincidences } \\
\hline
 \ar -\ar{} & prompt & 4$\cdot 10^{-6}$ \\
 \ar -\ar{} & late & 3$\cdot 10^{-4}$ \\
 \ar -Cherenkov & prompt & $\leq 4\cdot10^{-7}$ \\
 \ar -Cherenkov & late & $\leq$ 1$\cdot 10^{-4}$\ \\
 \hline
 \multicolumn{3}{l}{Correlated events } \\
 \hline
  Gamma-Cherenkov & prompt & $\leq 6\cdot10^{-5}$\\
  \hline
\end{tabular}
\end{center}
\label{tab:pileup}
\end{table}%

The pile-up cuts described in Sect.~\ref{subs:dataselection} were designed specifically to remove these events \cite{McElroy:2018vd}. 
Based on MC, Tab.~\ref{tab:pileup} gives an estimate of the fraction of the data that still contains pile-up after pile-up cuts. Fractions are given separately for the second event occurring within the prompt and the late \fp{} window. Coincidences within the prompt window tend to make the event more nuclear-recoil-like, while those in the late window tend to make the event look less like a nuclear recoil. 

\ar{} events that have either another \ar{} event or Cherenkov light randomly piled-up in the late time window are the most common coincidences. Their effect is small though, because the second event can have at most 10\% as much light as the first event for the pile-up cuts not to remove it.  For example, a 10~PE event (\ar{} or Cherenkov) piling up in the late part of a 120~PE \ar{} event with \fp=0.3 creates a pile-up event reconstructing at 130~PE and \fp=0.277. We verified through a toy MC that this level of contamination with pile-up in the late window does not affect the leakage predictions.

The next most common type of pile-up is correlated gamma-Cherenkov events. For example, 10 PE from Cherenkov light in prompt coincidence with a 120~PE ER event with \fp=0.3 create a pile-up event reconstructing at 130~PE and \fp=0.35. We simulated Cherenkov light production in PMT glass and acrylic as well as energy deposition in the LAr from \Nuc{Tl}{208} in the PMT glass using the full detector MC to obtain the event rate and PP distributions in the energy region of \SIrange{100}{200}{\totalnsc}. We then verified through a toy MC that the admixture of these events did not affect the leakage predictions for pure ER events, which this paper is concerned with. Based on the MC, the leakage of correlated pile-up events starts to dominate over that of pure ER events at a leakage probability for pure events smaller than $10^{-13}$.

\subsubsection{Non-ER events, blinding, and energy-dependence of the shape parameters}
The cuts used do not remove all backgrounds. For example, alpha events with degraded energy reconstruct just above the \ar-population in the PP. This leads to a small increase in the number of events at the upper edge of the \ar{} population. At the same time, data blinding removes some events from this region. 

The data blinding is implemented in low-level variables that do not correspond exactly to the PPs and energy estimator used here, hence there is no sharp line above which blinding affects the data in the variables shown here. The blinding removes fewer than 10 events per \totalnsc{} from the upper edge of the PP distributions.

To study the effect that events at the upper edge of the \ar{} PP distribution have on the fit, all fits were repeated with an upper fit limit in the PP parameter set to be approximately the last bin that still has more than 10 events. Furthermore, these fits were done using the 1D model,  Eq.~\eqref{eq:gamma} convolved with a Gaussian only, to additionally test the difference between the two modelling approaches. Fig.~\ref{fig:systematics} shows leakage predictions for \fpsc{} based on the standard fit with the 2D model, based on the fit with the full fit range in \fpsc{} using the 1D model, and based on the fit with restricted fit range in  \fpsc{} using the 1D model.

\subsubsection{Modelling of the NR signal region}
\begin{figure}[htbp]
  \centering
  \includegraphics[width=\linewidth]{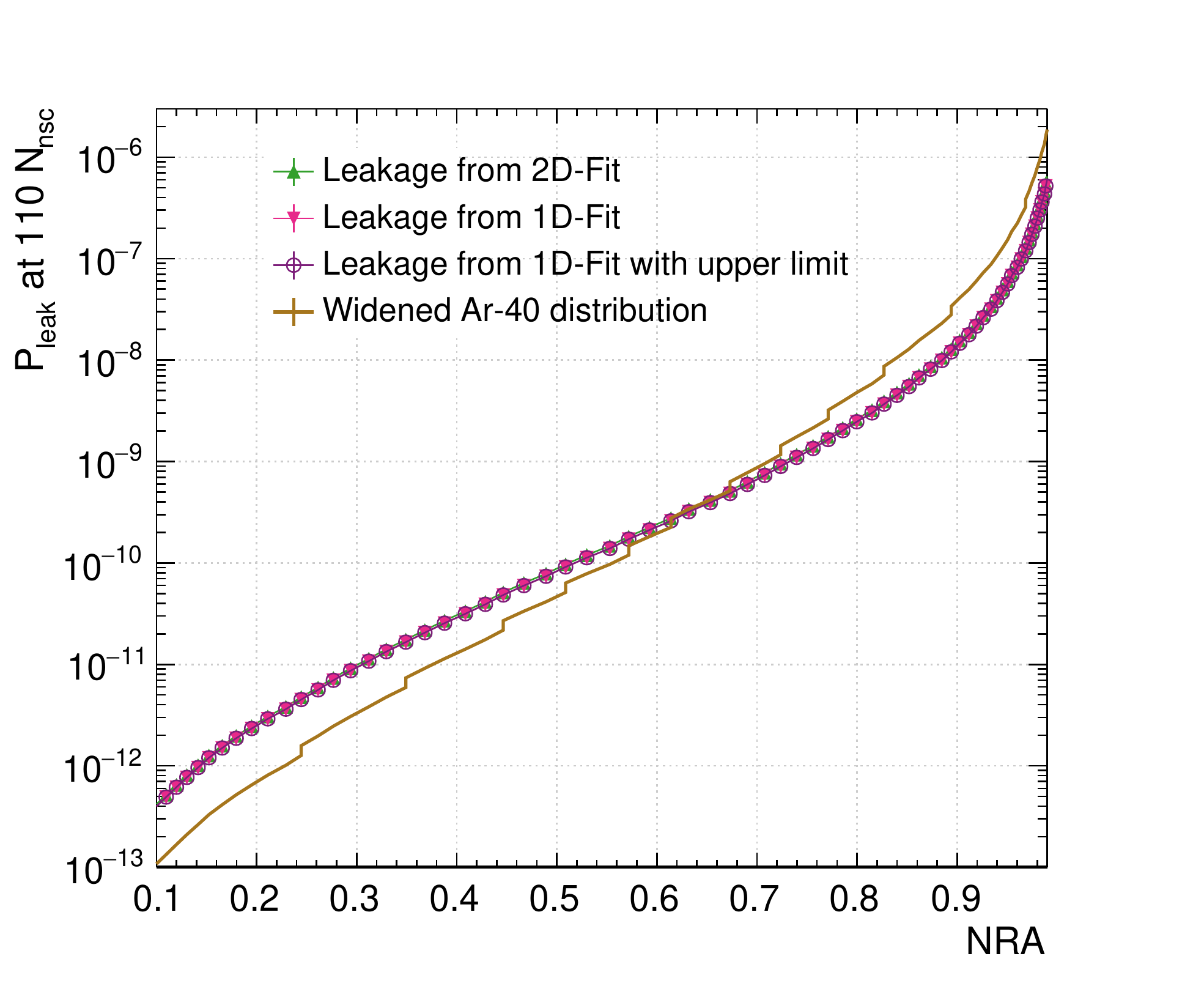}
  \caption{A study of three effects that lead to systematic uncertainties on the leakage probabilities. The green up-triangles show the nominal leakage probability for \fpsc{} from the 2D fit (the curve is the same as in Fig.~\ref{fig:leakagevse}). The pink down-triangles use the results from the 1D fit instead. The purple boxes are obtained with the 1D fit where the upper fit limit was reduced from 0.6 to 0.55 to exclude bins with fewer than 10 events. The points overlap, indicating that the fit procedure does not significantly change the leakage predictions. The brown curve uses an \arr{} distribution modified to account for energy resolution differences between MC and real data.}
    \label{fig:systematics}
\end{figure}

The position of the dark-matter signal region in each PP has systematic uncertainties related to how well the MC describes the detector and to incomplete knowledge of the ratio of singlet to triplet states created as a function of energy for nuclear recoil events. We consider here only the possible mis-modelling of the detector energy resolution, which has a significant effect on the width of the \arr{} distribution and a slight effect on its mean. For the \fp{} parameters, we construct alternate \arr{} event distributions with a larger energy resolution consistent with the width of the \ar{} PP distribution, as described in \cite{Ajaj:2019wi}. The leakage predictions based on this modified signal region are shown in Fig.~\ref{fig:systematics}.

\subsubsection{Data cleaning cuts}
After applying low-level and timing cuts, and the pile-up cut that requires that there is no more than one event peak within the digitization window, varying the cut values for the remaining cuts in reasonable ranges does not change the shape of the PP distribution in any significant way. 

\section{Stability of the discrimination power in time}\label{sec:stability}
\begin{figure}[htbp]
\centerline{\includegraphics[width=\linewidth]{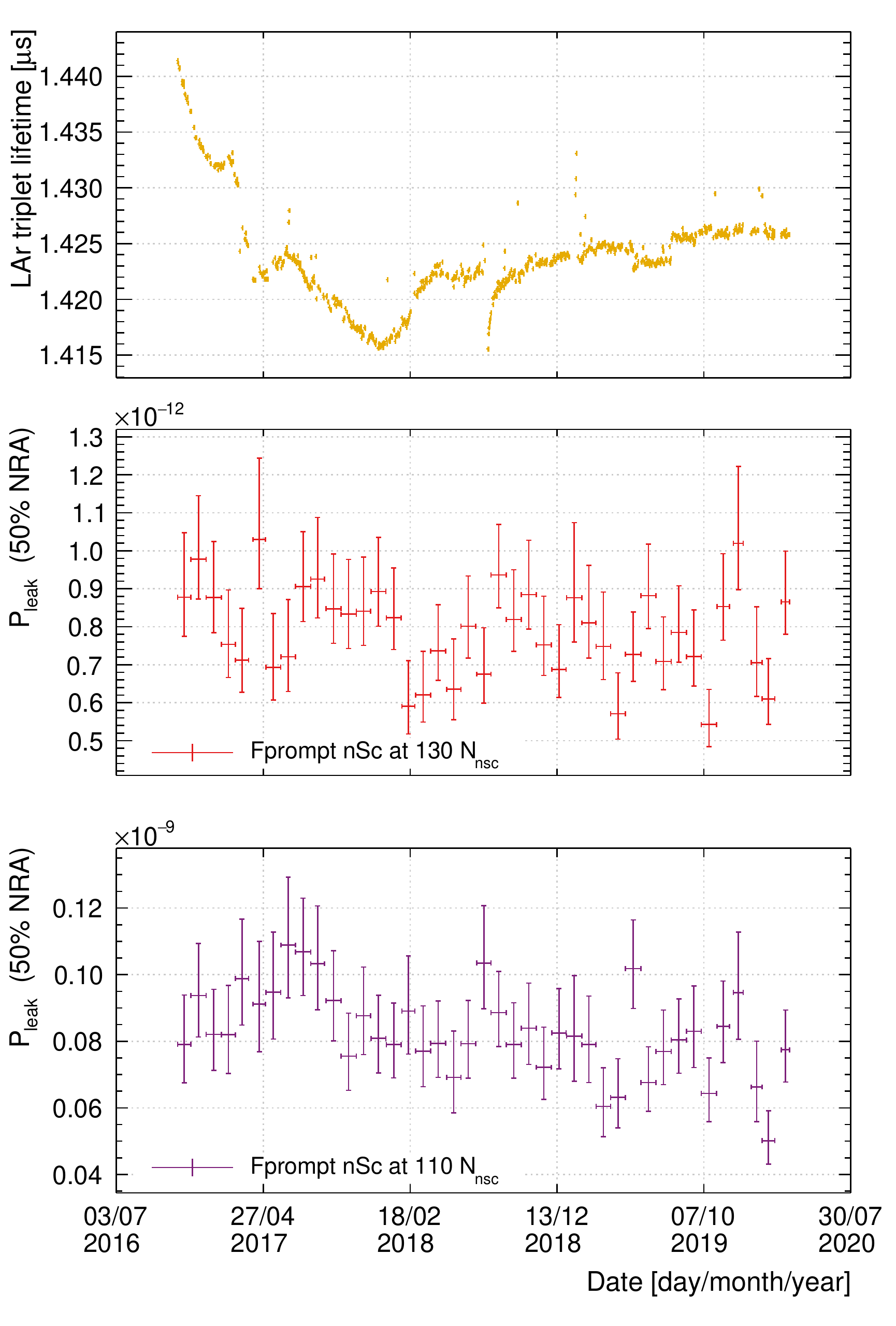}} 
\caption{The upper two figures show the time evolution of the leakage probabilities at \SI{110}{\totalnsc} and \SI{130}{\totalnsc} for \fpsc{} at 50\% NRA. The triplet lifetimes in the third figure are determined by using the full mathematical model of the \ar{} scintillation pulseshape \cite{DEAPCollaboration:2020hx}. The observed triplet lifetime strongly depends on the state variables of the detector, e.g. temperature, pressure, or impurities diffused from the detector into the liquid argon. A change in the triplet lifetime is expected to influence the shape of the PP distribution and hence causes the PP distribution of the whole data set to be a superposition of distributions with slightly different parameters.}
\label{fig:leakagetrend}
\end{figure}

The data shown here cover approximately 3.5~years of real time. We determined the leakage predictions individually per month to check if there were any changes due to detector degradation. Calibrations show no significant degradation of the light detection system, so the most significant effects are expected from changes in light yield or in the observed LAr triplet excimer lifetime due to impurities in the LAr. The triplet lifetime, which is sensitive to electro-negative impurities \cite{Acciarri:2009dj,Acciarri:2010gm}, is shown in Fig.~\ref{fig:leakagetrend} for each day when physics data were recorded. The triplet lifetime is obtained from fits to the \ar-pulseshape of events in the energy region considered here  \cite{DEAPCollaboration:2020hx}. The change over 3.5~years is less than 2\%. For the \fp{} PPs, this corresponds to a shift in the position of the PP distribution mean for the \ar{} population of approximately 0.8\%.

The leakage predictions shown in Fig.~\ref{fig:leakagetrend} are calculated against a fixed nuclear recoil band, that is the position of the NRA lines was not adjusted for changes in triplet lifetime. We see no significant degradation in PSD performance over the operation time of the DEAP-3600 detector.

\section{Discussion}\label{sec:photoninformation}

We consider the PSD parameter distribution for approximately \eventnumber{} events from \SIrange{16}{33}{keV_{ee}}, most of which represent electromagnetic interactions, largely \ar{} $\beta$-decays. The data contain a small fraction of pile-up events that do not significantly alter the distribution shape. Fig.~\ref{fig:residuals2d} shows that the effective fit model for the PP-distribution shape of ER events describes the data to better than 10\% accuracy over several orders of magnitude. Starting near the upper edge of the population of electromagnetic events, the model no longer describes the data well. In this region, non-ER backgrounds appear, and the event count is further biased by data blinding. This region does not significantly influence the model fit, as shown in Fig.~\ref{fig:systematics}

Based on the model fit, we calculate the probability for ER backgrounds to leak into the NR signal region. The latter is determined using Monte Carlo simulation. The simulation does not perfectly describe the shape of the PP distributions. For the \fpsc{} PP, an NR distribution shape is constructed to match measured shapes. The leakage probabilities shown in the remaining figures are systematically high or low by an amount as shown in Fig.~\ref{fig:systematics}, but the relative difference between the performance of the PPs remains the same.

Fig.~\ref{fig:leakagevse} shows that \lrqpe{} is an improvement over \fpqpe{}, but \lrsc{} provides little to no improvement over \fpsc{}. The fact that the \lr{} algorithm barely improves on \fp{} if the input is not significantly biased, even though \lr{} uses the full information available about photon arrival times, while \fp{} only considers whether or not photons arrived in the prompt window, can be understood by considering the weight functions.

The weight functions for the likelihood-based PP (Eq.~\ref{eq:lrecoilweights}) can be seen as the information carried by a photon as a function of the photon arrival time. A weight of $+1$ means the photon points toward the NR-hypothesis, while a weight of $-1$ means the photon points toward the ER-hypothesis. A weight of $0$ means that a photon detected at this time carries no information about the interaction type.

To compare the photon weight functions for $\fp$ and  $\lr$, the weight function for $\fp$ is manually constructed as a step function: It is $1$ in the prompt window (all photons here count toward the NR hypothesis), then $-1$ until the end of the total integration time (all photons here count toward the ER hypothesis), and $0$ at all later times (since they are not considered).

\begin{figure}[htbp]
\centerline{\includegraphics[width=\columnwidth]{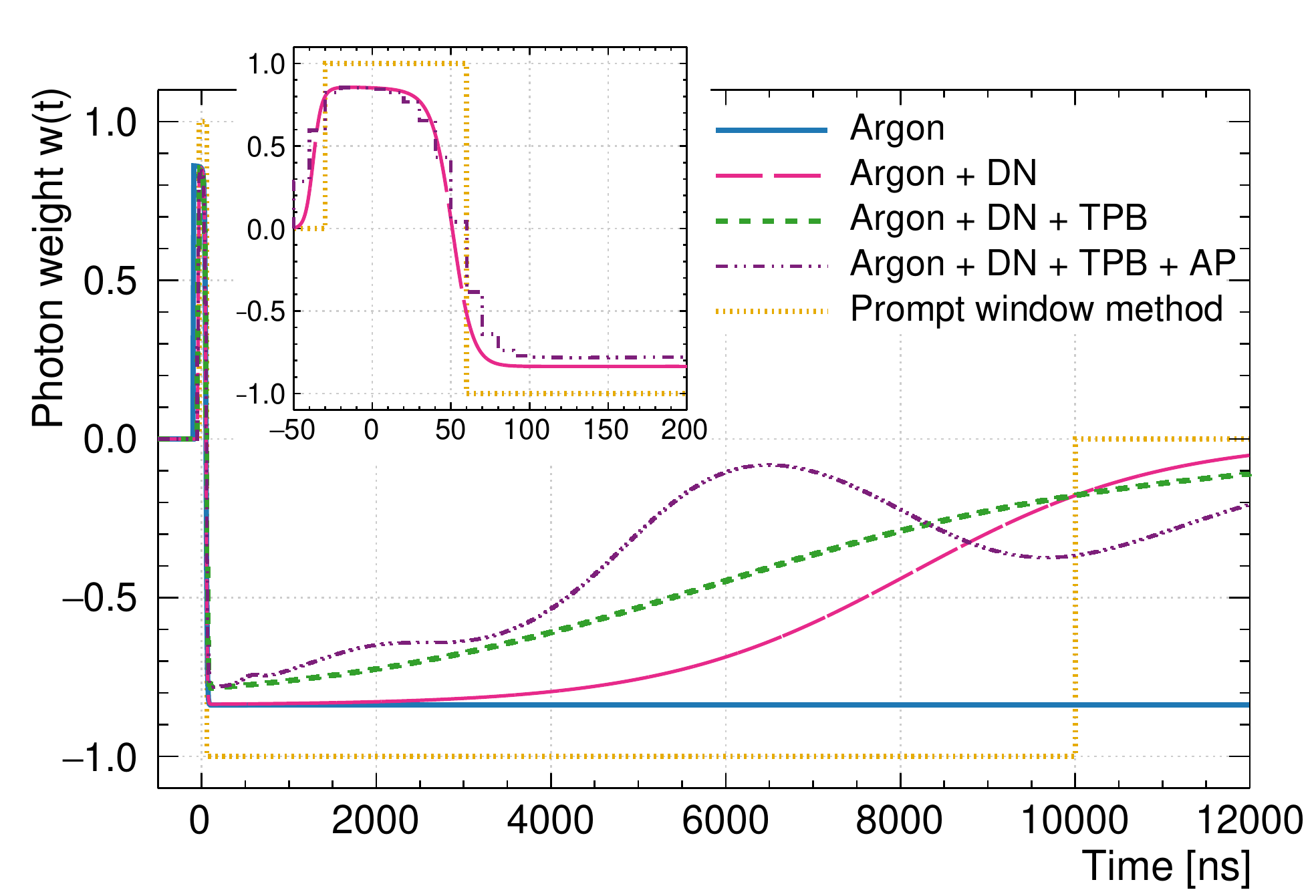}} 
\caption{The \lr{} weight-functions from Eq.~\eqref{eq:lrecoilweights} calculated for photon detection PDFs that take into account different components of the detector response  are shown. The blue solid line considers only the LAr scintillation (Argon), the pink long-dashed line adds dark noise (DN), the green short-dash line includes the slow component of TPB fluorescence (TPB), and the purple dot-dash line adds afterpulsing (AP). This is compared to the equivalent weight function used in the $\fp$ PP (yellow dotted line). The insert zooms in on the time region from \SIrange{-50}{200}{\nano\second}; for reasons of legibility, only three of the lines are shown here. Note that the detector's time resolution is included in the PDFs the photon weights are based on.  The weight can be interpreted as follows: PE detected at times where $w(t) > 0$ strengthen the NR hypothesis, while PE detected where $w(t) < 0$ strengthen the ER hypothesis. }
\label{fig:dvtime_model}
\end{figure}

The weight functions for $\fp$ and for $\lr$ are shown together in Fig.~\ref{fig:dvtime_model}. The $\lr$ weight function is shown for PDFs that include different detector effects, to illustrate their impact. The weights are always close to $+1$ near $t=0$, where most of the singlet light is detected. In the absence of detector effects, the weight function falls sharply after the singlet peak and stays constant afterwards, because at later times, the singlet component becomes negligible and the photon weight becomes the log-ratio of triplet fractions for ER and NR events. This shape is very similar to the step-function that is used for $\fp$. This means that the $\fp$ parameter already captures nearly all the available information.

When adding delayed TPB fluorescence and dark noise to the PDFs, the weight function still drops sharply after the singlet peak but then slowly rises towards zero as the signal-to-noise ratio becomes worse. Including AP in the model adds bumps to the weight function at the times of high AP probability; due to the AP, photons at this time contribute little information, which \lrqpe{} accounts for but \fpqpe{} does not.

\lrqpe{} cannot reliably reach the performance of \lrsc{} because \lrqpe{} accounts for AP based on the detected photon times over all PMTs.  The PE counting algorithm used to determine \nsc{} is applied individually for each PMT, where an AP must follow a previous pulse in the same PMT, so it uses more of the available information to make a better assessment as to whether or not a pulse is an AP.

\section{Conclusion}\label{sec:conclusion}

We presented how approximately \eventnumber{} \ar{} beta decay events between approximately \SIrange{16}{33}{\kilo\electronvolt_{ee}}, collected by the DEAP-3600 detector in \livedays~live-days, look under four different pulseshape discrimination methods, and predict their leakage probabilities into the nuclear-recoil signal region as a function of energy and nuclear recoil acceptance. With a light yield of \lynsc~\nsc/keV$_{ee}$, in a \SI{0.165}{\kilo\electronvolt_{ee}} wide bin at \SI{16}{\kilo\electronvolt} and allowing for a 50\% NRA, we estimate that with the \fpqpe{} PSD parameter a leakage of $7.5\cdot 10^{-9}$ is achieved, while \lrqpe{} reaches $2.3\cdot 10^{-9}$, and \fpsc{} and \lrsc{} both reach approximately $1\cdot 10^{-9}$. 

We find that due to the time structure of LAr scintillation, likelihood-based methods only improve on the prompt-fraction method if there is significant instrumental bias, in this case PMT afterpulsing, and information on this bias is included in the likelihood analysis. Otherwise, the prompt-fraction method captures nearly all the available information, leaving little room for the likelihood analysis to make a better assessment as to the particle type.

\section*{Acknowledgements}

We thank the Natural Sciences and Engineering Research Council of Canada,
the Canadian Foundation for Innovation (CFI),
the Ontario Ministry of Research and Innovation (MRI), 
and Alberta Advanced Education and Technology (ASRIP),
Queen's University,
the University of Alberta,
Carleton University,
the Canada First Research Excellence Fund,
the Arthur B.~McDonald Canadian Astroparticle Research Institute,
DGAPA-UNAM (PAPIIT No.~IN108020) and Consejo Nacional de Ciencia y Tecnolog\'ia (CONACyT, Mexico, Grant A1-S-8960),
the European Research Council Project (ERC StG 279980),
the UK Science and Technology Facilities Council (STFC) (ST/K002570/1 and ST/R002908/1), 
the Russian Science Foundation (Grant No.~16-12-10369), 
the Leverhulme Trust (ECF-20130496),
the Spanish Ministry of Science and Innovation (PID2019-109374GB-I00), 
and the International Research Agenda Programme AstroCeNT (MAB/2018/7)
funded by the Foundation for Polish Science (FNP) from the European Regional Development Fund.
Studentship support from
the Rutherford Appleton Laboratory Particle Physics Division,
STFC and SEPNet PhD is acknowledged.
We thank SNOLAB and its staff for support through underground space, logistical, and technical services.
SNOLAB operations are supported by the CFI
and Province of Ontario MRI,
with underground access provided by Vale at the Creighton mine site.
We thank Vale for their continuing support, including the work of shipping the acrylic vessel underground.
We gratefully acknowledge the support of Compute Canada,
Calcul Qu\'ebec,
the Centre for Advanced Computing at Queen's University,
and the Computation Centre for Particle and Astrophysics (C2PAP) at the Leibniz Supercomputer Centre (LRZ)
for providing the computing resources required to undertake this work.

\bibliography{bibliography}
\bibliographystyle{tp_unsrt_doi}

\end{document}